# Molecular Electronics: From Single-Molecule to Large-Area Devices.

Dominique Vuillaume

Institut for Electronics Microelectronics and Nanotechnology (IEMN)

CNRS, Univ. Lille, France.

Dominique.vuillaume@iemn.fr

**Abstract:** This mini review focuses on conductance measurements through molecular junctions containing few tens of molecules, which are fabricated along two approaches: (i) conducting atomic force microscope contacting a self-assembled monolayers on metal surface, and (ii) tiny molecular junctions made of metal nanodot (diameter < 10 nm), covered by fewer than 100 molecules and contacted by a conducting atomic force microscope. In particular, this latter approach has allowed to obtain new results or to revisit previous ones, which are reviewed here: (i) how the electron transport properties of molecular junctions are modified by mechanical constraint, (ii) the role of intermolecular interactions on the shape of conductance histograms of molecular junctions, and (iii) the demonstration that a molecular diode can operate in the microwave regime up to 18 GHz

## Introduction

Precise measurements of the electronic transport (ET) properties through molecular junctions are required for the development of molecular electronics. However, it is a challenging objective since several mechanisms can be involved. For example, the conductance through molecule junctions is very sensitive to atomic details of the molecule/electrode contact geometry, molecular conformation, as well as molecule-molecule interactions.[1-3] As a consequence, large number (hundreds or thousands) of conductance measurements and robust statistical analysis are mandatory. As reviewed in Refs [4] and [5], mechanically controllable break junctions (MCBJ) or scanning tunneling microscope MCBJ (STM-MCBJ) are methods of choice for the measurement of the conductance of single (or a few) molecules. Several groups have reported the existence of several peaks in the histograms of conductances of molecular junctions made of alkyl molecules between gold electrodes.[6-14] These peaks in the conductance histograms[8] were attributed to several mechanisms, for example, different atomistic configurations of the molecule on the electrode surface (i.e. molecule attached to a gold ad-atom, or grafted on a hollow site of the gold surface).[8, 11] Other mechanisms are the variability of orientation of the molecules with respect of the normal of the surface,[15] as well as local variations of the crystallographic orientations of the Au electrode.[16] However, literature also reports discrepancies in these experiments which may come from variability of the experimental conditions (e.g. these experiments were performed in a liquid environment with various solvents, the speed of operation of the MCBJ or STM-MCBJ were different, different data filtering methods,…). For instance, other groups reported the observation of a single peak, or three peaks or even no clear peak in the conductance histograms of alkylthiol junctions.[14, 17-19]

At the macroscopic level (see a review in Ref [20]), it is also possible to construct conductance histograms from molecular junctions made with GaIn eutectic drop, Hg drop, or lithographed electrodes. However, these measurements are more demanding, or time consuming, and these features result in a smaller number of counts in the histograms.[21-28] In these cases, the measured conductance histograms usually displayed only a single peak because the large size of the electrodes (few $\mu m^2$ to $mm^2$) results in an averaging of the microscopic features.

This mini review focuses in between these two situations, at the mesoscopic scale, and reports on conductance measurements through molecular junctions fabricated along two approaches: (i) conducting atomic force microscope contacting a self-assembled monolayers on metal surface, and (ii) tiny molecular junctions made of metal nanodot (diameter < 10 nm), covered by fewer than 100 molecules and contacted by a conducting atomic force microscope.

## Conducting atomic force microscopy on self-assembled monolayers.

Conducting-atomic force microscope (C-AFM) on self-assembled monolayers (SAM) is the technique usually used to address the electron transport properties though a few hundred molecules, but many works reported only average values of the conductance. Nevertheless, a few groups measured conductance histograms by C-AFM, and reported a single peak of conductance whatever the molecules (alkylthiols, molecular switches) and at different measurements conditions (e.g. loading force).[27, 29-30]

Among the pioneering works on C-AFM in molecular electronics, Wold and Frisbie examined the electron transport through alkylthiols SAMs on Au by C-AFM and reported how the measured current is dependent on the C-AFM tip loading force.[31-32] They observed a force threshold (around 100 nN) above which the current dramatically increases due to the fact that mechanical indentation becomes dominant and induces severe deformation in the SAM (see more detail below in section "*Effect of mechanical*

*strain*"). At low loading force, they observed an electron tunnel mechanism characterized by a tunnel decay constant, β, of ca. 1 Å$^{-1}$ in good agreement with data from macroscopic junctions (see Refs [20,1]). Conversely, increasing the loading force allows to study the role of disorder on ET and several groups explained the increase of the current (when increasing loading force) by an intermolecular enhancement of electron transfer with an increase of the molecule tilt angle [33-34] (induced by the tip loading force) and/or variation of the interface dipole at the interface between the molecules and the metal electrodes.[35] C-AFM was also used to study ET through SAMs of conjugated molecules. Again, this approach was proven consistent (working a low loading forces to avoid detrimental effect of mechanical indentation) with already know data with β ~ 0.4 Å$^{-1}$ for oligophenylene[36] and oligothiophene.[37] C-AFM technique was also used to study ET through "long" (up to 7 nm) π-conjugated oligomers ("molecular wires") - a series of oligophenyleneimine.[38] They revealed a transition between tunneling transport (with β ~ 0.3 Å$^{-1}$) to a hopping mechanism for molecules longer than ca. 4 nm, for which the ET is less length-dependent ("effective β" ~ 0.09 Å$^{-1}$). We note that this behavior resembles the one observed for organic thin films (oligothiophene, 4-22 nm thick),[39] albeit a third regime (for films between 8-16 nm) was also observed in this latter case.

## Nanodot molecule junctions

Recently, a novel approach was proposed to measure easily the conductance histograms of a great number of molecular junctions (up to $10^6$) from a single C-AFM image (figure 1).[40] This approach is based on the use of a large array of tiny (<10 nm) nanodot molecule junctions (NMJ), fabricated (e-beam lithography) on single crystal gold nano-electrodes, embedded in a degenerated (highly-doped) Si back contact, Fig. 1a.[41] Fewer than one hundred molecules are chemically grafted on each nanodot. During a scanning C-AFM image, each time the C-AFM tip passes over an individual NMJ, the current (at a given voltage) is directly recorded (the silicon substrate between NMJ is passivated by a thin silicon dioxide layer, and thus insulating and no measurable current are recorded by the C-AFM tip between NMJs). Typically, in a few µm$^2$ C-AFM image (Fig. 1c), 2000-4000 NMJs are measured in a single scan. Current or conductance histograms are simply extracted from these C-AFM images and repeating this approach at various bias allow the measurement of 2D histograms of the current-voltages (I-V) curves.[40] With this NMJ approach, focusing on the archetype alkylthiol junctions, it was observed that the number of the conductance peaks depends on the atomic details of the electrode surface: 2 peaks for NMJs on single crystal Au nanodot vs. 3 peaks for SAMs on more disordered polycrystalline Au surfaces.[40] In the case of NMJs and by varying the chain length (from 8 to 18 carbon atoms) of the alkylthiols, the two conductance peaks (Fig. 1d) were ascribed to different phases of molecular organization, with the HC (high conductance) peak corresponding to a more ordered molecular organization in the junction in agreement with an early hypothesis[42] that the "all-trans" conformation of the alkyl chains in the SAMs (ordered and closely compact SAMs) results in a more efficient electron transport than when the molecules have "gauche" defects (as in disordered SAMs).

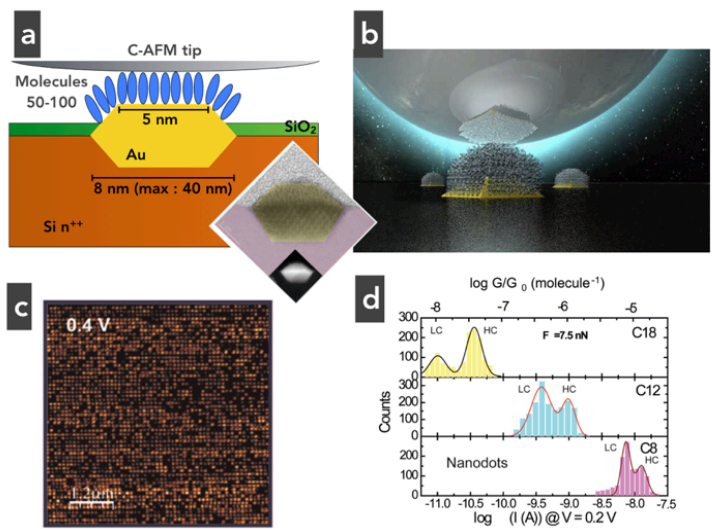

**Figure 1. (a)** Scheme of the cross-section of the NMJ (nanodot-molecule-junction), **(b)** artist view of the NMJ. The inset shows a high resolution TEM image of the gold nanodot, the scale bar (white) is 5 nm (TEM inset reprinted with permission from Ref. [40], copyright (2011) Wiley). **(c)** Typical current image (at 0.4 V) measured by C-AFM on a network of ~ 1600 NMJs with alkylthiol molecules. **(d)** Current histograms for alkylthiol NMJs with n=8, 12 and 18 carbon atoms at 0.2 V. Figs. 1c and 1d reprinted with permission from Ref. [41], copyright (2012) American Chemical Society).

**Effect of mechanical strain**

C-AFM on alkylthiol SAMs on Au surface, but with a large lateral extension, was used to study the effect of a mechanical constraint, the loading force of the C-AFM tip, on the ET properties of these molecular junctions. However, these previous results showed puzzling and contradictory results. Some groups reported that the tunnel decay constant, β, is almost constant when increasing the C-AFM loading force,[42-43] while other groups have observed a slightly increasing[44] or decreasing[32] β. The NMJ approach is ideal for revisiting the mechanical strain effect on ET properties of molecular junctions[45] because (i) the loading force can be finely adjusted; (ii) the curvature radius of the C-AFM tip (~40 nm) is larger than the nanodot electrodes (<10 nm) and the metal/molecules/metal device has an ideal parallel plate structure with no dependence of the molecular junction area on the loading force and (iii) a statistical analysis (conductance histograms) can easily be performed as discussed above.

The main result reported in Ref. [45] revealed unprecedented behaviors for NMJs of alkylthiol monolayers (8-18 carbon atoms chain length). In this case, an increase of the loading force (from 3 to 30 nN) induced a large decrease of the tunnel decay factor β from 0.9 to 0.2/C atom (Fig. 2). A spectroscopic analysis of the current-voltage (IV) curves (by the transient voltage spectroscopy, TVS, method[46-51]) showed that the HOMO level (with respect to the Au electrode Fermi energy) decreased by ~0.4 eV. These features were ascribed to a modification of the dipole at the Au/alkylthiol interface induced by the applied force on the molecules. These results were supported by DFT calculations.[35,45,52] Such a large modulation of β and HOMO level, induced by the applied force, were not reported in previous C-AFM measurements on SAMs deposited on Au substrate electrodes with a large lateral size (Refs. [32], [42-44]) and this feature can be related to several reasons: (i) in these previous C-AFM, the contact area increases with the loading force (indentation of the SAM) which complicates the interpretation of the measurements; (ii) in these previous experiments, SAMs were grafted on evaporated Au with a polycrystalline structure, and the defects/disorder at the interface between Au and alkylthiol may have hidden the behavior reported in Ref. [45].

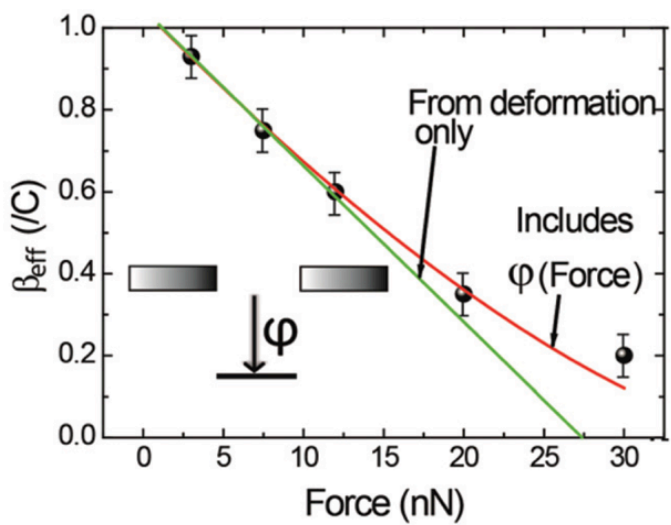

**Figure 2.** Figure Caption. Effective tunnel current decay factor measured from a series of alkylthiol NMJs (8, 12 and 18 carbon atoms) versus the C-AFM loading force. The green line shows a model taken only into account the mechanical deformation and the red line is for a model combining mechanical deformation and force-induced modification of the electronic structure of the NMJ. Reproduced from Ref. [45] with permission from The Royal Society of Chemistry.

**Estimation of the π-π intermolecular interaction energy**

The π-π intermolecular interactions have a key effect in the ET properties of devices based on π-conjugated materials, such as self-assembled monolayer field-effect transistors,[53-57] π-stacked organic nanowires and nanostructures,[58-60] and, generally speaking, in any organic electronic devices (OFET, OLED, OPV).[61] The probability of charge transfer between adjacent molecules when their π orbitals overlap is described by a key parameter, the transfer integral t, which can be estimated by quantum-chemical calculations[62] and photoelectron spectroscopy.[63] However, it is also highly desirable to determine the parameter t directly from the electronic properties of the molecular/organic devices. For example, in a metal/molecules/metal junction, due to this cooperative effect between molecules, it is known that the conductance of N identical molecules in parallel is not systematically equal to N time the conductance of the isolated molecule.[64-65] Recently, Reuter *et al.* suggested that such cooperative effects induce asymmetrical conductance histogram spectra (i.e. no longer log-normal distributed), and they proposed a simple toy model with two π-conjugated molecules in parallel between electrodes.[66] An experimental proof requires several conditions : (i) an easy way to record conductance histograms (a solid statistical analysis requires thousands of molecular junctions), (ii) a method to control/monitor the intermolecular interactions. The NMJ approach is well adapted, because the number of molecules in the junctions and the degree of intermolecular interactions can be varied by changing the size of the bottom nanodot electrode (10 - 40 nm in diameter) and by making well-compact monolayers of π-conjugated molecules or diluted ones (e.g. with alkyl chains) to increase the intermolecular distance. This approach was used with NMJs of ferrocenylalkylthiol ($FcC_{11}SH$).

The results of these experiments[67] showed that the energy of the π-π intermolecular interaction can be deduced from both electrochemistry (which gives the coupling between the charge distributions on adjacent molecules) and molecular electronics (which gives the coupling between adjacent molecular orbitals) by doing cyclic voltammetry (CV) and conductance histogram measurements on the same large array of ferrocenylalkylthiol NMJs (about 3000 NMJs). From CV measurements, dense monolayers in the NMJs give a broadening of the voltammograms (FWHM=110 mV) as expected for intermolecular electrostatic coupling, while less dense, diluted, monolayers in the NMJs have a narrow CV peak (FWHM=40 mV). Similarly, conductance histograms for the dense monolayers clearly exhibited an asymmetric distribution with a tail towards the low conductance values (Fig. 3a) as predicted by Reuter *et al.*,[66] while a classical log-normal distribution was observed for diluted monolayers (Fig. 3b). Ultra-High-Vacuum Scanning Tunneling Microscopy (UHV-STM) was used to resolve the supramolecular organization of the dense monolayer (fig. 3c), which was used as the model for DFT calculations of t. To fit the conductance histograms, the previously proposed model with 2 molecules[66] was extended to a network of 11x11 molecules more relevant considering the size of the NMJs (about 150 molecules). Good fits (fig. 3d) of the asymmetric histograms were obtained with t ~ 30-35 meV, in very good agreement with the calculated DFT values for these NMJs.

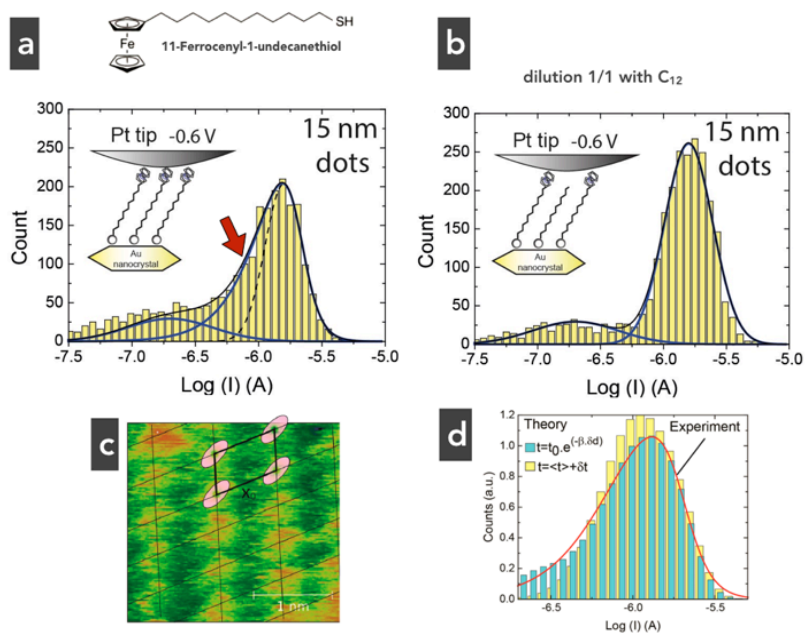

**Figure 3.** Current histograms of ferrocenylalkylthiol NMJs (~ 3000 NMJs) at -0.6 V and a loading force of 15 nN for **(a)** a densely packed monolayer to favor π-π intermolecular interaction and **(b)** for a diluted monolayer with alkylthiol to cancel the intermolecular interactions. **(c)** High-resolution UHV-STM image of the molecules (orange spots) on the top surface of the Au nanodot showing the unit cell organization on the surface with an average area per molecule of 0.40 $nm^2$. **(d)** Simulated histograms fitting the experimental results (red line) with an intermolecular energy interaction of ~ 30-35 meV. Reprinted with permission from Ref. [67], copyright (2017) American Chemical Society.

**High-frequency molecular electronics**

Up to now, molecular electronics devices were demonstrated and characterized in the DC regime, or at frequencies below MHz. In a device perspective, this feature represents a drastic limitation. However, theoretical studies predicted transit times through molecular junctions as fast as few femtoseconds,[68-69] allowing, in principle, operations in the THz regime. Molecular diodes are archetype molecular devices since the pioneering work of Aviram and Ratner.[70] Metal/ferrocenylalkylthiol (FcC$_{11}$SH)/metal junction is a typical example of a molecular diode, which was deeply studied in DC regime[71-72] and used in a half-wave rectifier bridge working at 50 Hz.[73] An interesting question is to know what is the upper limit of frequency at which such a molecular diode can operate without a degradation of the rectifying properties.

Several issues have to be considered to operate a molecular device at high frequencies. A first issue is the need of tiny devices to reduce the device capacitance since, in first approximation, the cut-off frequency is $f_C = G/(2\pi C)$, G being the conductance and C the capacitance of the device. Basically, a single molecule device fulfills this condition, albeit the conductance cannot exceed the quantum of conductance $G_0$=77.5 µS, and the measured conductances are in the range of $10^{-3}$-$10^{-2}$ $G_0$ for single π-conjugated molecule junctions. Consequently, one need to design a molecular junction considering an optimum number of molecules in parallel in order to increase G of the junction as much as possible, while keeping C small enough. The NMJ approach is a good compromise with reported G of about 0.36 mS for Au/ferrocenylalkylthiol (FcC$_{11}$SH)/PtIr junction.[74] The second issue is that high frequency (GHz and above) measurements usually require devices matching a 50Ω impedance, far below the "high-impedance" of molecular devices. This limitation can be overcome using a recently developed interferometric scanning microwave microscope (iSMM)[75-76] combining a scanning tunneling microscope (STM) with AC (alternative current) microwave signal (up to 20 GHz) superimposed on the tip. Note that similar approaches (DC + AC STM around few GHz) were reported and successfully used to compare the polarizabilities of various molecular junctions[77-78] or to measure the spin resonance of a terbium-phthalocyanine derivative (at 5K)[79-80] but none of these works reported a direct measurement of the dynamic conductance in this microwave frequency regime and at room temperature.

Using a large array of NMJs (few tens of nanometers in size with about 150 ferrocenylalkylthiol molecules) with the iSMM, DC current and microwave (up to 17.8 GHz) properties were simultaneously measured (Fig. 4).[74] The measured microwave reflexion signal S$_{11}$ clearly showed a diode rectification ratio of 12 dB (Figs. 1b and 1c), as the DC current (Fig. 1a). The dynamic conductance vs. voltage curves G$_{HF}$-V extracted from the S$_{11}$ measurements are strictly superimposed to DC conductance-voltage curve (Figs. 1d and 1e) demonstrating that the molecular rectification behavior is preserved up to 17.8 GHz. In addition to a high dynamic conductance of G$_{HF}$=0.36 mS, a total capacitance (intrinsic molecular junction capacitance and fringe capacitance due to tip-substrate coupling) of ca. 110 aF was determined. These values allowed an estimation of $f_C$ = 520 GHz for these molecular diodes, on a par with RF-silicon Schottky diodes.[74]

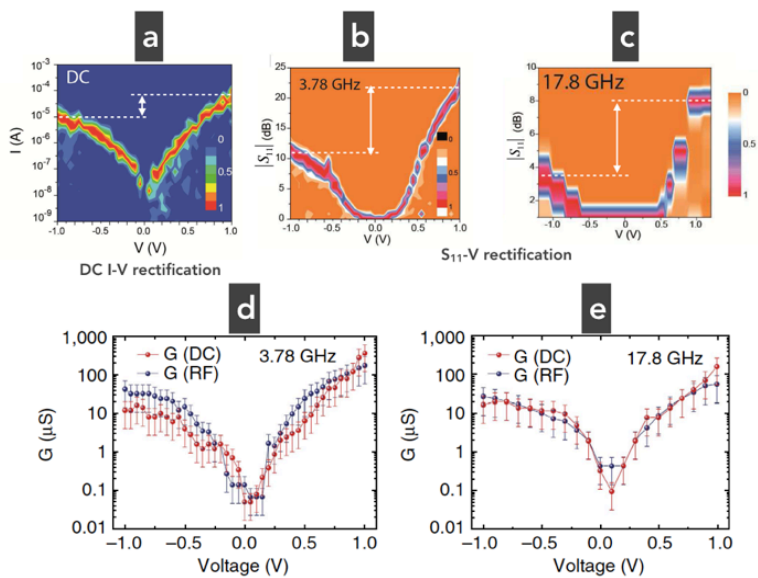

**Figure 4.** 2D histograms of **(a)** DC current-voltage curve , **(b)** S$_{11}$-voltage curve at 3.78 GHz, **(c)** S$_{11}$-voltage curve at 17.8 GHz measured simultaneously on the same network of ferrocenylalkylthiol NMJs (at loading force of 18 nN). **(d)** and **(e)** : comparison of the hf conductance (extracted from the S$_{11}$-V curves) and the dc conductance versus voltage for hf measurements at 3.78 GHz and 17.8 GHz, respectively. Reprinted from Ref. [74], Creative Commons Attribution 4.0 International License.

## Summary and outlook

This mini review has discussed the electron transport properties of molecular junctions with size in between single molecule devices and large area molecular junctions. In particular, several results were reviewed, which may have impacts in several applications or connected research fields. The effect of mechanical strain on the electron transport properties of molecular junctions, and especially the observed very great change in the tunneling rates, could be useful to develop pressure or force sensors at the molecular scale. The shape of the conductance histograms of molecular junctions is an efficient technique to determine the interaction energy (transfer integral) between adjacent molecules, which is known as one of the key parameters to understand and optimize the performances of more macroscopic organic devices (e.g. the charge mobility in organic transistors). Molecular devices are also prone for high-frequency operations and the demonstration of a molecular diode at 18 GHz opens the door to operate organic and molecular devices in the high-frequency regime. This results will allow exploring theoretically predicted new effects, such as, dynamic resonances inducing an amplification of the conductance of the molecular devices,[81-82] or, in future practical applications the use of organic diodes in energy harvesting systems[83] (at Wi-Fi and other electromagnetic radiation) on flexible substrates. We also note that some unexpected results not reported previously (e.g. the large force load dependency of the tunnel factor and HOMO position, the "two-peaked" distribution of conductance on nanodot-molecule junctions) ought to be compared with results from other molecular device platform for a better understanding and towards a global consensus on electron transport through molecular nanostructures.

## Acknowledgements

I acknowledge all colleagues at IEMN who have carried out the work reported in this minireview: N. Clément, who has initiated and supervised the work on NMJs, J. Trasobares, F. Vaurette, K. Smaali, D. Théron, with the collaborations of J. Rech, C. Walh, T. Jonckeere, T. Martin (Centre de Physique Théorique, CNRS, Univ. Marseille, France), O. Aleveque, E. Levillain (Moltech Anjou, CNRS, Univ. Angers, France), V. Diez-Cabanes, Y. Olivier, J. Cornil, P. Leclère (Laboratory for Chemistry of Novel Materials, University of Mons, Belgium), G. Fotti, T. Frederiksen, D. Sanchez-Portal, A. Arnau (Donostia International Physics Center, Donostia-San Sebastián, Spain), G. Patriarche (C2N, CNRS, Univ. Orsay, France), who contributed to different parts of this work. Financial supports: EU-FP7 Nano-microwave project, Renatech (the French national nanofabrication network) and Equipex Excelsior project

**Keywords:** electron transport • molecular electronics • monolayers • nanotechnology • scanning probe microscopy